\documentclass[conference]{IEEEtran}
\IEEEoverridecommandlockouts

\usepackage{cite}
\usepackage{amsmath,amssymb,amsfonts}
\usepackage{graphicx}
\usepackage{textcomp}
\usepackage{xcolor}
\usepackage{bm}
\usepackage{booktabs}
\graphicspath{{figures/}}

\usepackage{tikz}
\usetikzlibrary{positioning, fit, calc, backgrounds}

\newcommand{\uS}{\,\mu\mathrm{S}}
\newcommand{\rev}{\color{blue}}
\newcommand{\plotwidth}{0.9\columnwidth}

\begin{document}

\title{Achievable Rate and Reconstruction Fidelity of FeFET-Based Compute-in-Pixel Arrays
}

\author{\IEEEauthorblockN{Mohamed Akrout$^1$, Md Rahatul Islam Udoy$^1$, Ahmedullah Aziz$^1$, Nuria Gonzalez-Prelcic$^2$, Robert W. Heath$^2$}
\IEEEauthorblockA{$^1$\textit{EECS Department}, \textit{University of Tennessee}, Knoxville, USA, \\
$^2$\textit{ECE Department}, \textit{University of California San Diego}, San Diego, USA, \\
\{makrout,aziz\}@utk.edu, mudoy@vols.utk.edu, \{ngprelcic,rwheathjr\}@ucsd.edu}
}
\maketitle
\begin{abstract}
Compute-in-pixel (CiP) image sensors such as MACPix weight each photodiode voltage by a conductance stored in a ferroelectric FET and sum the currents on a column bus, so that $N$ pixels are read as $M<N$ weighted sums before digitization. We model the column bus as a Gaussian multiple-access channel and derive its achievable rate under thermal and shot noise, device variation, and ADC distortion. Conductance variation acts as noise proportional to the scene energy, which limits the rate to $4.5$ bits per readout and makes a 6-bit ADC sufficient. End-to-end simulations recover natural images at $26.3$\,dB from 2 bits per pixel and, for sparse scenes, reduce the edge-node energy by up to $6.4\times$ against raw 8-bit readout.
\end{abstract}


\section{Introduction}
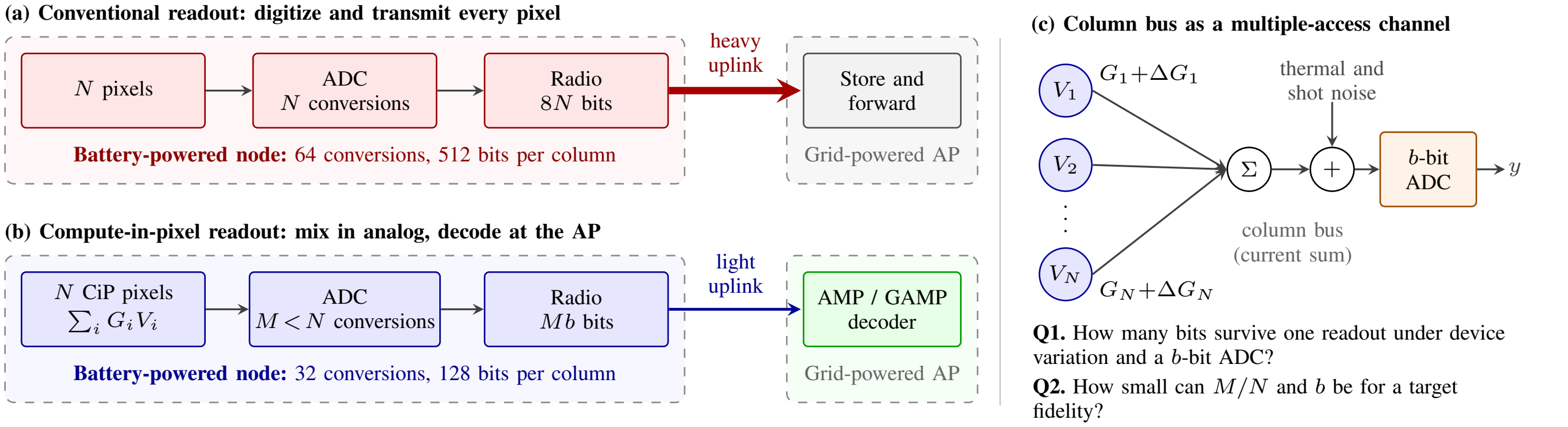
\begin{figure*}[!t]
\centering
\begin{tikzpicture}[x=1mm, y=1mm, >=stealth, font=\scriptsize,
  blk/.style={draw=#1!60!black, fill=#1!10, rounded corners=1.5pt, minimum height=8.5mm,
              minimum width=21mm, align=center, inner sep=1.5pt, line width=0.5pt},
  grp/.style={draw=black!45, dashed, rounded corners=3pt, inner sep=1.8mm, line width=0.5pt},
  ttl/.style={font=\scriptsize\bfseries, anchor=south west, inner sep=0pt},
  note/.style={font=\scriptsize, inner sep=0.5pt, align=center},
  flow/.style={->, line width=0.6pt, black!75}]
\node[blk=red] (a1) at (11, 0) {$N$ pixels};
\node[blk=red] (a2) at (37.5, 0) {ADC\\$N$ conversions};
\node[blk=red] (a3) at (64, 0) {Radio\\$8N$ bits};
\draw[flow] (a1) -- (a2); \draw[flow] (a2) -- (a3);
\node[note, text=red!55!black] (an) at (37.5, -7.5) {\textbf{Battery-powered node:} 64 conversions, 512 bits per column};
\node[blk=gray, minimum width=18mm] (aap) at (99, 0) {Store and\\forward};
\node[note, text=black!60] (aapn) at (99, -7.5) {Grid-powered AP};
\begin{scope}[on background layer]
  \node[grp, fill=red!3, fit=(a1)(a3)(an)] (A) {};
  \node[grp, fill=black!3, fit=(aap)(aapn)] (AAP) {};
\end{scope}
\draw[->, line width=2.4pt, red!65!black, shorten >=0.6pt] (a3.east) -- (aap.west)
  node[midway, above=1mm, note, text=red!55!black] {heavy\\uplink};
\node[ttl] at ($(A.north west) + (0, 1.3mm)$) {(a) Conventional readout: digitize and transmit every pixel};
\node[blk=blue] (b1) at (11, -25) {$N$ CiP pixels\\[1pt]$\textstyle\sum_i G_i V_i$};
\node[blk=blue] (b2) at (37.5, -25) {ADC\\$M\!<\!N$ conversions};
\node[blk=blue] (b3) at (64, -25) {Radio\\$Mb$ bits};
\draw[flow] (b1) -- (b2); \draw[flow] (b2) -- (b3);
\node[note, text=blue!55!black] (bn) at (37.5, -32.5) {\textbf{Battery-powered node:} 32 conversions, 128 bits per column};
\node[blk=green, minimum width=18mm] (bap) at (99, -25) {AMP / GAMP\\decoder};
\node[note, text=black!60] (bapn) at (99, -32.5) {Grid-powered AP};
\begin{scope}[on background layer]
  \node[grp, fill=blue!3, fit=(b1)(b3)(bn)] (B) {};
  \node[grp, fill=green!4, fit=(bap)(bapn)] (BAP) {};
\end{scope}
\draw[->, line width=0.9pt, blue!60!black, shorten >=0.6pt] (b3.east) -- (bap.west)
  node[midway, above=1mm, note, text=blue!55!black] {light\\uplink};
\node[ttl] at ($(B.north west) + (0, 1.3mm)$) {(b) Compute-in-pixel readout: mix in analog, decode at the AP};
\draw[black!25, line width=0.4pt] (112.5, 6) -- (112.5, -36);
\node[ttl] at (116, 6.3) {(c) Column bus as a multiple-access channel};
\foreach \y/\lab in {0/1, -8.5/2, -21/N}{
  \node[draw=blue!60!black, fill=blue!10, circle, minimum size=6mm, inner sep=0pt, line width=0.5pt] (u\lab) at (120, \y) {$V_{\lab}$};
}
\node at (120, -13.6) {$\vdots$};
\node[draw, circle, minimum size=5mm, inner sep=0pt, line width=0.5pt] (sum) at (141, -9) {$\Sigma$};
\draw[flow] (u1.east) -- (sum.west);
\draw[flow] (u2.east) -- (sum.west);
\draw[flow] (uN.east) -- (sum.west);
\node[note, anchor=south west] at ($(u1.east) + (0.5, 0.6)$) {$G_1{+}\Delta G_1$};
\node[note, anchor=north west] at ($(uN.east) + (0.5, -0.6)$) {$G_N{+}\Delta G_N$};
\node[draw, circle, minimum size=5mm, inner sep=0pt, line width=0.5pt] (add) at (150.5, -9) {$+$};
\node[blk=orange, minimum width=11mm] (adc) at (161.5, -9) {$b$-bit\\ADC};
\draw[flow] (sum) -- (add); \draw[flow] (add) -- (adc);
\draw[flow] (adc.east) -- ++(3.2mm, 0) node[right, inner sep=0.8pt] {$y$};
\node[note, text=black!70] (noise) at (150.5, 1.2) {thermal and\\shot noise};
\draw[flow] (noise.south) -- (add.north);
\node[note, text=black!60] at (146, -17.5) {column bus\\(current sum)};
\node[note, align=left, text width=57mm, anchor=north west] at (116, -26.5)
  {\hyphenpenalty=10000 \textbf{Q1.} How many bits survive one readout under device variation and a $b$-bit ADC?\\[1.5pt]
   \textbf{Q2.} How small can $M/N$ and $b$ be for a target \mbox{fidelity}?};
\end{tikzpicture}
\vspace{-0.3cm}
\caption{Motivation. (a) A conventional node digitizes and transmits every pixel. (b) A compute-in-pixel (CiP) array forms $M<N$ weighted current sums in the analog domain and leaves the reconstruction to the grid-powered access point (AP); numbers are for $N=64$ (Section~\ref{sec:results}). (c) Pixels as users of a multiple-access channel whose gains are FeFET conductances perturbed by variation $\Delta G_i$, followed by a $b$-bit ADC.}
\label{fig:motivation}
\end{figure*}

Visual sensors at the network edge spend most of their energy on digitizing pixels and transmitting them. A conventional node converts all $N$ pixels of a column and sends the full payload over its radio, as shown in Fig.~\ref{fig:motivation}(a). The link is asymmetric: the node runs on a battery, whereas the access point (AP) is grid-powered and can afford heavy computation, as in massive machine-type communications \cite{liu_yu}.

In-sensor computing moves processing into the pixel array, before digitization \cite{wan_insensor,wang_vision,jang_insensor,yang_dynamic}. The MACPix pixel \cite{mac} combines a photodiode with a ferroelectric field-effect transistor (FeFET) whose multi-domain polarization stores a non-volatile analog weight, so that each pixel multiplies its illumination by the weight and the pixels of a column add their currents on a shared line. Such \emph{compute-in-pixel} (CiP) arrays can replace $N$ conversions by $M<N$ weighted sums (see Fig.~\ref{fig:motivation}(b)), acting as a compressive imager \cite{duarte_spc,oike_cs} without a separate modulator.

At first glance, transmitting weighted sums should not be cheaper than transmitting pixels, since a weighted-sum readout requires more ADC bits than a per-pixel readout. The saving comes from needing fewer readouts. Real scenes have fewer degrees of freedom than pixels: sparse scenes are mostly dark, and neighboring pixels of natural images are similar \cite{candes_wakin}. A conventional node can exploit this redundancy only by digitizing every pixel and compressing digitally, whereas every CiP sum carries information about every pixel of the column, so the compression happens in the analog domain for a few picojoules and the AP solves the reconstruction problem. Since the radio spends $0.1$ to $0.4\,\mu$J per bit \cite{siekkinen}, the energy of the node is essentially proportional to the bits it sends.

What remains unquantified is (i) how much information the analog weighted sums retain, and (ii) how far the number of weighted sums $M$ and the ADC resolution can be reduced. Both questions have counterparts in communication theory as depicted in Fig.~\ref{fig:motivation}(c). Pixels that drive one bus can be seen as users of a multiple-access channel (MAC), and unmixing them at the AP is a compressed-sensing problem \cite{donoho_cs} which can be solved with approximate message passing (AMP) \cite{donoho_amp} and generalized AMP (GAMP) \cite{rangan_gamp}, as in massive-access receivers.

Adopting this communication-theoretic perspective for the CiP readout and the AP processing, we derive the achievable rate of the column bus under thermal and shot noise, conductance variation, and ADC distortion. We also show that, to an uncalibrated receiver, conductance variation acts as noise proportional to the scene energy, and we design decoders that use the exact likelihood of the quantizer. Finally, we provide an end-to-end evaluation from scene to reconstructed image, together with an estimate of the energy saved at the node.

\section{FeFET-Based Compute-in-Pixel Array Architecture}
\label{sec:architecture}

Our proposed communication framework builds upon the MACPix architecture \cite{mac}, which integrates light sensing, non-volatile weight storage, and analog multiply-and-accumulate operations within an image sensor array.

\subsection{Pixel Array Organization and Circuit Configuration}

\begin{figure*}[!t]
    \centering
    \includegraphics[width=0.97\textwidth]{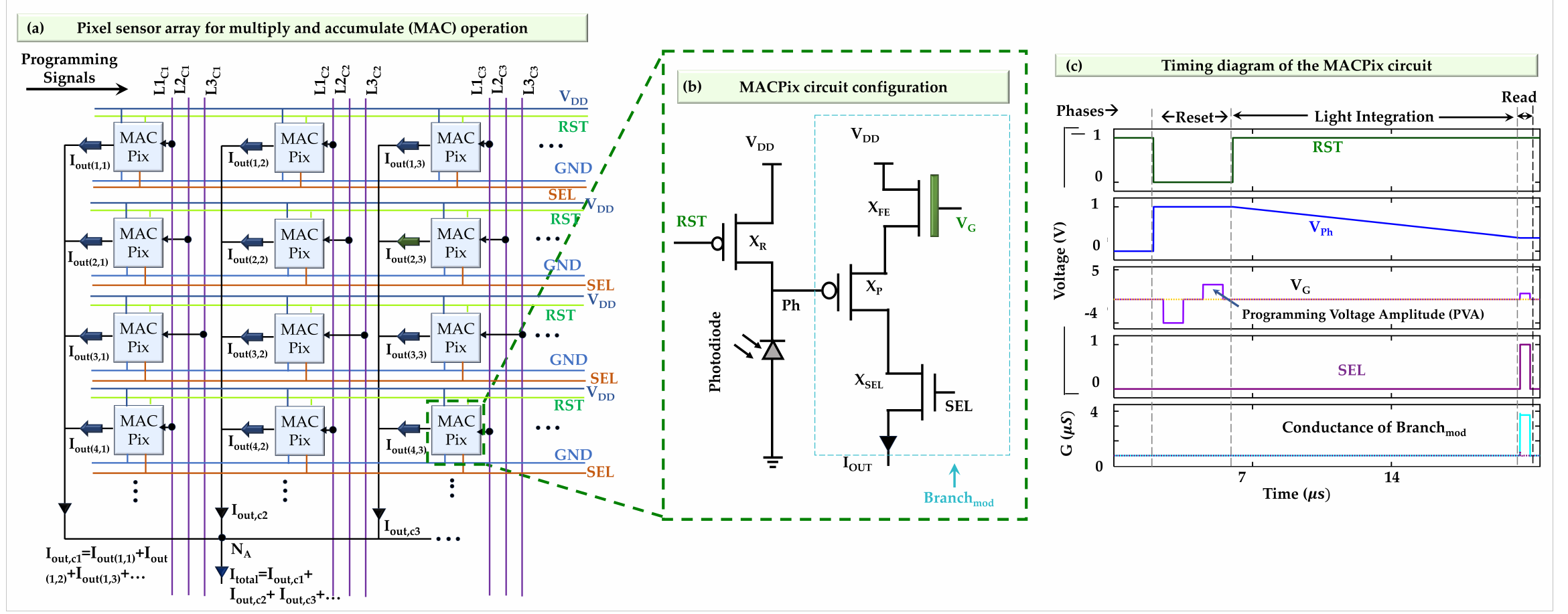}
    \caption{MACPix sensor array architecture and pixel circuit configuration. (a) Organization of the MACPix pixel array with shared control lines and column-wise current accumulation, (b) circuit configuration of an individual MACPix cell incorporating a photodiode and programmable FeFET. (c) Operating sequence of the MACPix circuit, including the reset, light integration, and readout phases{\rev .}}
    \label{fig:macpix_array}
    \vspace{-0.3cm}
\end{figure*}

Fig.~\ref{fig:macpix_array} illustrates the MACPix array organization and the underlying pixel circuit. As shown in Fig.~\ref{fig:macpix_array}(a), the image sensor consists of a two-dimensional array of MACPix pixels connected through shared row and column lines. Each row shares the supply voltage ($V_{\mathrm{DD}}$), reset (RST), selection (SEL), and ground signals. Each column incorporates three FeFET gate control lines, denoted as $L_1$, $L_2$, and $L_3$, to independently control the programmable weights and activation of pixels at different row positions. Pixels at three-row intervals share the same line. This supports independent weights for the three rows of a $3\times3$ kernel. The output currents of pixels within each column are connected to a common readout line, enabling their individual contributions to be accumulated directly in the analog domain.

The circuit configuration of an individual MACPix cell is presented in Fig.~\ref{fig:macpix_array}(b). The circuit integrates a photodiode with a reset transistor ($X_R$), a PMOS transistor ($X_P$), a pixel selection transistor ($X_{\mathrm{SEL}}$), and an n-type ferroelectric field-effect transistor ($X_{\mathrm{FE}}$). The FeFET introduces programmable conductance into the readout branch, enabling the pixel output current to depend on both the sensed illumination and the stored weight.

By controlling the programming voltage amplitude (PVA), different polarization levels can be established, providing non-volatile analog weight storage. The FeFET operates as a source degeneration element for $X_P$, and the combined conduction path, denoted as $\mathrm{Branch}_{\mathrm{mod}}$, determines the output current during readout.

\subsection{Pixel Operation and Analog Multiplication}

The operating sequence of the MACPix circuit is illustrated in Fig.~\ref{fig:macpix_array}(c). Each sensing cycle consists of reset, light integration, and readout phases. During the reset phase, $X_R$ is activated to initialize the photodiode node to $V_{\mathrm{DD}}$. When a new weight is required, the FeFET is programmed through a sequence of gate voltage pulses. Since the state is non-volatile, programming is repeated only when the weight changes.

During the light integration phase, the reset transistor is turned off, and the photodiode generates photocurrent in response to incident illumination. The accumulated photocharge causes the photodiode node voltage to decrease from its initialized value. For a fixed integration time ($T_{\mathrm{int}}$) and photodiode capacitance, the voltage drop ($\Delta V_{\mathrm{Ph}}$) is proportional to the photocurrent and represents the optical input to the pixel.

During readout, the pixel selection transistor and FeFET are activated, allowing the circuit to generate an output current ($I_{\mathrm{out}}$) that depends on both the photodiode voltage and the stored conductance. In other words, the pixel implements an approximate analog multiplication between the illumination-dependent input and the stored weight. For the communication model, we denote the photodiode voltage drop of pixel $i$ by $V_i=\Delta V_{\mathrm{Ph},i}$ and define $G_i$ as its effective programmed multiplication coefficient. The pixel output is approximated as
\begin{equation}
I_{\mathrm{out},i} \approx G_i V_i,
\label{eq:pixel_mult}
\end{equation}
where $G_i$ represents the effective input-to-output coefficient of the complete pixel readout branch rather than solely the isolated FeFET conductance. We neglect offsets and residual nonlinearity.

\begin{figure}[!t]
    \centering
    \includegraphics[width=0.85\columnwidth]{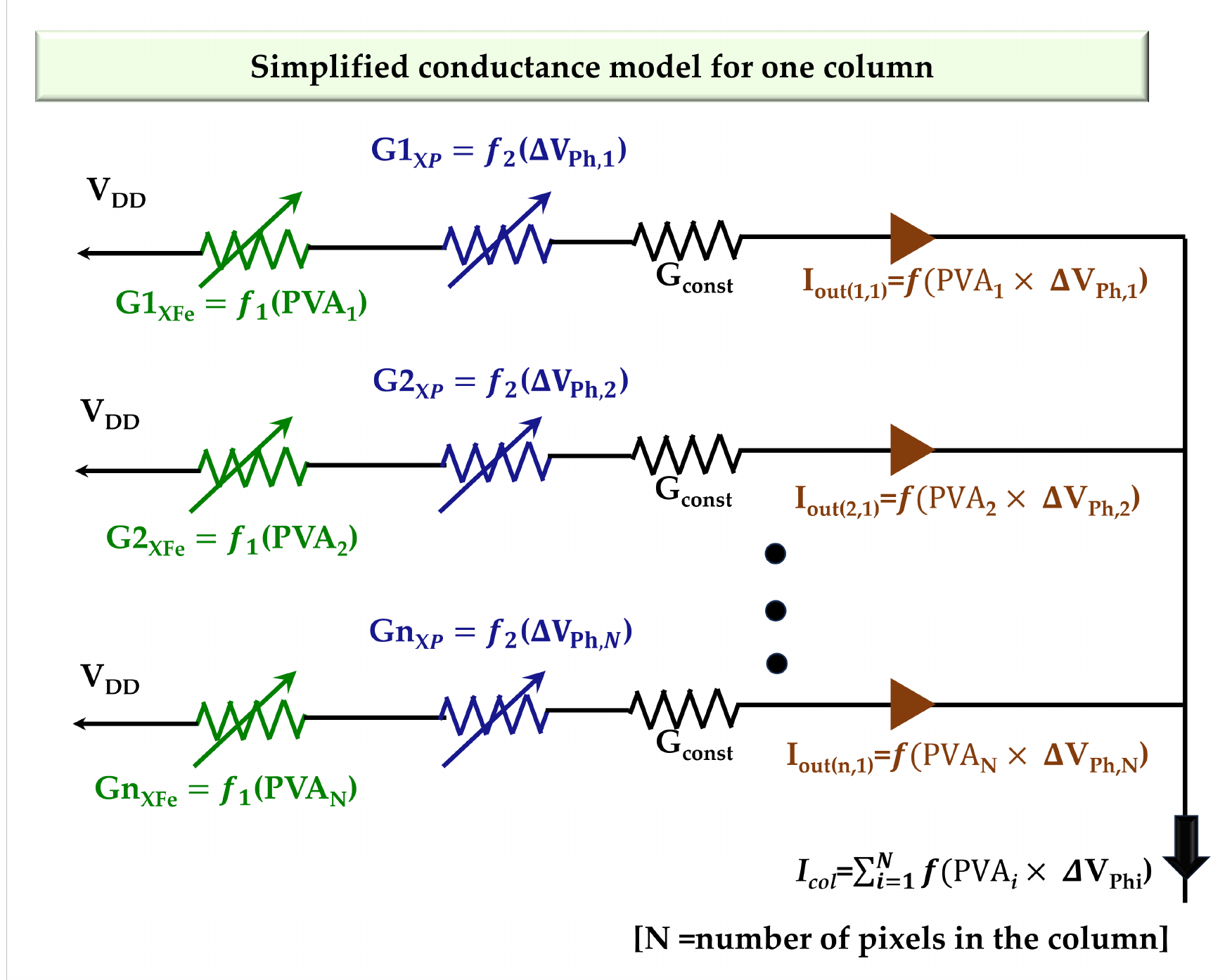}
    \caption{Simplified conductance model of a MACPix column, illustrating the effective conductance contributions of individual pixels connected through the shared column readout line.}
    \label{fig:column_model}
    \vspace{-0.3cm}
\end{figure}

\subsection{Column-Parallel Accumulation and Analog Measurement Formation}

As illustrated in Fig.~\ref{fig:column_model}, each pixel contributes an output current determined by its illumination-dependent input and programmed conductance. When multiple pixels within a column are activated simultaneously, their currents are combined along the common readout line according to Kirchhoff's current law.

For a column containing $N$ activated pixels, substituting \eqref{eq:pixel_mult} gives the total column current
\begin{equation}
I_{\mathrm{col}} = \sum_{i=1}^{N} I_{\mathrm{out},i} \approx \sum_{i=1}^{N} G_i V_i.
\label{eq:column_mac}
\end{equation}

The shared column thus acts as a physical analog encoder that maps $N$ illumination-dependent inputs into one electrical measurement.

In the original MACPix architecture, the programmable weights implement convolution kernels. Here, they are regarded as elements of a measurement matrix: acquiring $M$ weighted measurements of $N$ pixels yields $\mathbf{G}\in\mathbb{R}^{M\times N}$, each row of which is one weighting configuration. This assumes that the scene is static over the measurement sequence and that the configurations can be programmed within the frame time. The superposition of pixel currents on the shared line is then analogous to signal superposition in a multiple-access channel, which Section~\ref{sec:mac} formalizes by adding the noise and device variation that limit a real readout.

\section{The Column Bus as a Multiple-Access Channel}
\label{sec:mac}
\subsection{Channel Model}
Superposition on the bus in \eqref{eq:column_mac} has the form of a MAC in which $N$ transmitters, the pixels, reach one receiver, the column ADC, through gains $G_i$ \cite{cover}. A physical readout differs from \eqref{eq:column_mac} in that the realized conductance deviates from its programmed value and the current is noisy:
\begin{equation}
    y = \sum_{i=1}^{N} (G_i+\Delta G_i)\, V_i + n .
\label{eq:mac}
\end{equation}
Here $\Delta G_i\sim\mathcal{N}(0,\sigma_G^2)$ models device-to-device variation and $n\sim\mathcal{N}(0,\sigma_n^2)$ collects thermal noise of the parallel cells and shot noise,
\begin{equation}
    \sigma_n^2 = \Big( \frac{4 k_B T}{R_{\mathrm{eq}}} + 2 q\,(I_{\mathrm{dc}}+\bar I_{\mathrm{col}}) \Big) B ,
\label{eq:noise}
\end{equation}
with $R_{\mathrm{eq}}=1/(N\bar G)$ the equivalent resistance for mean conductance $\bar G$, $I_{\mathrm{dc}}$ the dark current, $\bar I_{\mathrm{col}}$ the mean signal current, and $B=1/(2T_{\mathrm{int}})$ the bandwidth of one readout per integration period.

\subsection{Achievable Rate}
Let the pixel voltages be independent with variance $\sigma_V^2$ and second moment $\mathbb{E}[V^2]$, and let $\overline{G^2}$ denote the mean-square programmed conductance. A receiver that knows the programmed weights but not their deviations sees the useful power $P_S$ and the variation-induced interference $P_\Delta$,
\begin{equation}
    P_S = N\,\overline{G^2}\,\sigma_V^2, \qquad P_\Delta = N\,\sigma_G^2\,\mathbb{E}[V^2] .
\label{eq:powers}
\end{equation}
The ADC is a uniform $b$-bit quantizer with full scale $\pm\lambda\sigma_y$, where $\sigma_y^2=P_S+P_\Delta+\sigma_n^2$. Its distortion has a granular part of step $\Delta=2\lambda\sigma_y/2^b$ and an overload part due to clipping,
\begin{equation}
    P_Q = \frac{\Delta^2}{12} + 2\big[(1+\lambda^2)\,Q(\lambda)-\lambda\,\phi(\lambda)\big]\sigma_y^2 ,
\label{eq:quant}
\end{equation}
with $\phi$ and $Q$ the standard normal density and tail function. Since the aggregate of many pixels is close to Gaussian, treating the three impairments as additive noise gives the achievable rate
\begin{equation}
    R = \frac{1}{2} \log_2 \!\Big( 1 + \frac{P_S}{\sigma_n^2+P_\Delta+P_Q} \Big)\ \ \text{bits/readout},
\label{eq:rate}
\end{equation}
that is, $R/T_{\mathrm{int}}$ bits per second.

Two properties of \eqref{eq:rate} matter for design. First, the signal-to-interference ratio
\begin{equation}
    \frac{P_S}{P_\Delta} = \frac{\overline{G^2}}{\sigma_G^2}\cdot\frac{\sigma_V^2}{\mathbb{E}[V^2]}
\label{eq:sir}
\end{equation}
depends neither on the number of pixels per column $N$ nor on the integration time $T_{\mathrm{int}}$. Unlike thermal noise, variation cannot be averaged out by integrating longer or by activating more pixels, because it scales with the signal. Second, $P_Q$ is worth reducing only until it falls below $P_\Delta$, which fixes a useful ADC resolution.

The rate \eqref{eq:rate} bounds what a readout carries about the column as a whole, not whether individual pixels can be separated; that depends on the decoder, treated next.

\section{Recovering the Pixels at the Access Point}
\label{sec:amp}
\subsection{Linear Inverse Problem}
Stack the pixel voltages of a column in $\mathbf{x}=[V_1,\dots,V_N]^{\mathsf T}$ and the weights of readout $m$ in row $m$ of $\mathbf{G}\in\mathbb{R}^{M\times N}$, with entries drawn independently and uniformly from the programmable levels of mean $\bar G$ and standard deviation $\sigma_L$. The common-mode term $\bar G\,\mathbf{1}^{\mathsf T}\mathbf{x}$ is identical in all readouts and conveys only the column sum. It is measured once per column by a reference readout with all cells at the same level, digitized with 8 bits, and its code centers the window of the $b$-bit ADC for the $M$ mixed readouts, so that the converter spans only the informative fluctuation. After removal of the common mode and normalization by $\sigma_L\sqrt{M}$, $M$ uses of \eqref{eq:mac} read
\begin{equation}
    \mathbf{y} = \mathcal{Q}\big( (\mathbf{A}+\Delta\mathbf{A})\,\mathbf{x} + \mathbf{w} \big),
\label{eq:cs}
\end{equation}
where $A_{mi}=(G_{mi}-\bar G)/(\sigma_L\sqrt{M})$ has zero mean and variance $1/M$, $\Delta\mathbf{A}$ has i.i.d.\ $\mathcal{N}(0,\kappa^2/M)$ entries with $\kappa=\sigma_G/\sigma_L$, $\mathbf{w}\sim\mathcal{N}(\mathbf{0},\sigma_w^2\mathbf{I})$ is the normalized noise of \eqref{eq:noise}, and $\mathcal{Q}$ is the quantizer applied entrywise. The measurement ratio is $\delta=M/N$, and the AP knows $\mathbf{A}$ but not $\Delta\mathbf{A}$.

\subsection{Conductance Variation as Signal-Dependent Noise}
Conditioned on $\mathbf{x}$, the perturbation $\Delta\mathbf{A}\,\mathbf{x}$ is Gaussian with covariance $(\kappa^2\|\mathbf{x}\|^2/M)\,\mathbf{I}$ and is independent of $\mathbf{A}$ and $\mathbf{w}$. To a receiver that uses the nominal matrix, variation is therefore indistinguishable from additional white noise. Before quantization, \eqref{eq:cs} is equivalent to $\mathbf{A}\mathbf{x}+\mathbf{w}_{\mathrm{eff}}$ with
\begin{equation}
    \sigma_{\mathrm{eff}}^2 = \sigma_w^2 + \frac{\kappa^2}{\delta}\,\mathbb{E}[x^2] .
\label{eq:sigeff}
\end{equation}
Equation \eqref{eq:sigeff} is the reconstruction-side counterpart of \eqref{eq:sir}. The variation term grows with the scene energy and with $1/\delta$, so compressing harder raises the noise floor, and no amount of thermal signal-to-noise ratio removes it.

\subsection{Decoding Quantized Measurements}
A coarse quantizer is not well described by additive noise, so the decoder uses its exact likelihood. For quantizer thresholds $\{\tau_k\}$, $k$-th output level $q_k$, and $z_m=[\mathbf{A}\mathbf{x}]_m$,
\begin{equation}
    p(y_m = q_k \,|\, z_m) =  \int_{\tau_k}^{\tau_{k+1}} \mathcal{N}(u; z_m, \sigma_{\mathrm{eff}}^2)\, \mathrm{d}u ,
\label{eq:qchan}
\end{equation}
with the outermost cells extending to $\pm\infty$ so that clipping is modeled exactly \cite{kamilov}. Since $\sigma_{\mathrm{eff}}^2$ depends on $\|\mathbf{x}\|^2$ through \eqref{eq:sigeff}, it is re-estimated from the current estimate at every iteration. We combine \eqref{eq:qchan} with two priors. For scenes that are sparse in the pixel domain, GAMP \cite{rangan_gamp} with the minimum mean-square error denoiser of the prior takes \eqref{eq:qchan} as its output channel. Without quantization, it reduces to AMP \cite{donoho_amp}{\rev ,} whose error is predicted by a state evolution \cite{bayati} driven by \eqref{eq:sigeff}. Natural images are not sparse pixel by pixel, and for them the AP solves, jointly over all columns $\mathbf{x}_c$ of the image $\mathbf{X}$,
\begin{equation}
    \hat{\mathbf{X}} = \arg\min_{\mathbf{0}\le\mathbf{X}\le V_{\max}}\; -\sum_{c,m}\log p\big(y_{c,m}\,|\,\mathbf{a}_{c,m}^{\mathsf T}\mathbf{x}_c\big) + \mu\,\|\mathbf{X}\|_{\mathrm{TV}}.
\label{eq:map}
\end{equation}
Here, $\mathbf{a}_{c,m}^{\mathsf T}$ is row $m$ of the normalized matrix of column $c$, $\|\cdot\|_{\mathrm{TV}}$ denotes the total variation{\rev ,} and $\mu>0$ is a regularization weight, and \eqref{eq:map} is solved by accelerated proximal gradient iterations \cite{beck_tv}. In both decoders, the AP needs only the transmitted codes and the programmed weights.

\section{Numerical Results}
\label{sec:results}
\subsection{Setup}
Table~\ref{tab:params} lists the parameters of the simulated $64\times64$ array, which follow the device characterization of MACPix \cite{mac}. The MACPix readout branch provides eight effective conductance levels (3 bits), controlled through FeFET programming, which we approximate as uniformly spaced over the simulated range. The spread $\sigma_G=0.107\uS$ results from a $\pm9\%$ ferroelectric thickness tolerance and a $\pm150$\,mV threshold-voltage shift, giving $\kappa=0.05$. The device parameters were obtained from HSPICE simulations \cite{mac}. The ADC full scale is $\lambda=3$ standard deviations. For the rate, pixel voltages are uniform on $[0,1]$\,V. Reconstruction is evaluated on eight $64\times64$ test photographs and on synthetic scenes that are sparse in the pixel domain, in which each $V_i$ is zero with probability $1-\rho$ and uniform on $(0,1]$\,V otherwise, with $\rho=0.1$.

\begin{table}[!t]
\caption{Array Parameters}
\label{tab:params}
\centering
\footnotesize
\begin{tabular}{@{}ll@{}}
\toprule
Pixels per column, $N$ & 64 \\
Conductance levels & 8, from $1.16$ to $7.7\uS$ \\
Level mean $\bar G$, std.\ $\sigma_L$ & $4.43\uS$, $2.14\uS$ \\
Conductance variation, $\sigma_G$ & $0.107\uS$ ($\kappa=0.05$) \\
Integration time, bandwidth & $20\,\mu$s, $25$\,kHz \\
Equivalent resistance, $R_{\mathrm{eq}}$ & $3.53$\,k$\Omega$ \\
Dark current, $I_{\mathrm{dc}}$ & $0.0963\,\mu$A \\
\bottomrule
\end{tabular}
\end{table}

\begin{figure}[!t]
\centerline{\includegraphics[width=\plotwidth]{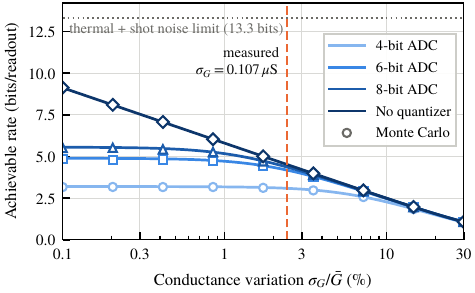}}
\caption{Achievable rate in \eqref{eq:rate} of one column bus versus conductance variation. Markers refer to Monte Carlo simulations of \eqref{eq:mac} with a saturating ADC.}
\label{fig:rate}
\vspace{-0.2cm}
\end{figure}

\begin{figure*}[!t]
\centering
\includegraphics[width=0.95\textwidth]{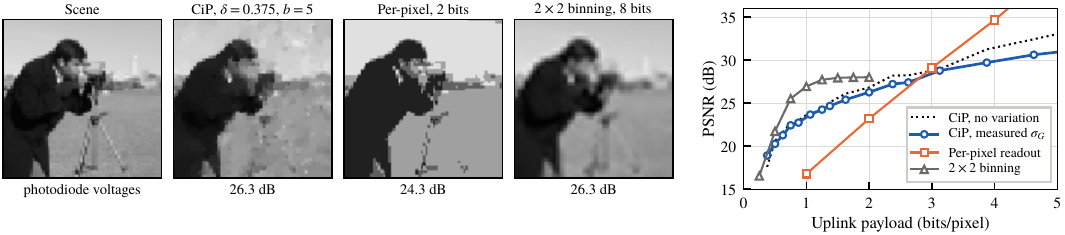}
\caption{End-to-end simulation of the $64\times64$ array with the measured conductance variation. Left: a scene and its reconstructions from 2 bits per pixel by CiP readout with random weights and by two conventional readouts. Right: PSNR averaged over eight test images versus uplink payload; each CiP marker is one pair $(\delta,b)$ on the upper envelope.}
\vspace{-0.4cm}
\label{fig:image}
\end{figure*}

\subsection{Rate of the Column Bus}
Fig.~\ref{fig:rate} plots \eqref{eq:rate} against the relative variation $\sigma_G/\bar G$. Markers are obtained by simulating \eqref{eq:mac} directly with random weights and a saturating quantizer, with the rate computed from the measured correlation between the weighted sum and the ADC output. They agree with the closed form to within $0.1$ bits. Thermal and shot noise alone would allow $13.3$ bits per readout. At the measured spread, the bus is deep in the variation-limited regime: the $27.2$\,dB of \eqref{eq:sir} lies $53$\,dB below the noise-limited SNR, and the rate falls to $4.52$ bits ($226$\,kb/s), corresponding to a loss of $66\%$. A 6-bit converter retains $92\%$ of the variation-limited rate and an 8-bit one $97\%$, whereas 4 bits retain $68\%$. Resolution beyond about 6 bits is thus spent on digitizing device mismatch, and only a smaller $\sigma_G$ raises the rate. For example, halving it adds one bit per readout.

In all simulations, each cell is subject to an independent deviation $\Delta G_i\sim\mathcal{N}(0,\sigma_G^2)$ that is drawn in every trial and unknown to the receiver. Without calibration, decoding uses the programmed weights and accounts only for the power of the deviation, through \eqref{eq:rate} and \eqref{eq:sigeff}, which is the worst case. If the static part of the deviation were measured once and reported to the AP, only the residual spread would remain and the operating point would move left along Fig.~\ref{fig:rate}: a residual of one half or one quarter of $\sigma_G$ gives $5.5$ or $6.5$ bits per readout without quantization, and $5.0$ or $5.4$ bits with an 8-bit ADC.

\subsection{End-to-End Image Reconstruction}
We pass images through the complete signal chain. Each photograph is mapped to photodiode voltages on $[0,1]$\,V, and every column is read with one reference readout followed by $M$ weighted-sum readouts in which the realized weights deviate from the programmed ones by fresh $\Delta G_i$ and the column current carries the thermal and shot noise of \eqref{eq:noise}. A saturating $b$-bit ADC with a fixed window of $\pm29.7\,\mu$A around the reference code produces the transmitted codes, so that the payload is $\delta b+8/N$ bits per pixel. The AP reconstructs the image from the codes and the weights with \eqref{eq:map}, where $\mu$ is tuned for each operating point on two further images.

Fig.~\ref{fig:image} shows a reconstruction and the peak signal-to-noise ratio (PSNR) averaged over the test images. With 5 bits and $\delta=0.375$, that is, 2 bits per pixel or a quarter of the raw 8-bit payload, CiP reaches $26.3$\,dB; $1.06$ and $3.1$ bits per pixel give $23.7$ and $28.8$\,dB. Conductance variation costs less than $0.6$\,dB up to 2 bits per pixel and $2.2$\,dB at 5: at low payload the error is set by the number of readouts and by the ADC, and the variation term of \eqref{eq:sigeff} limits only the high-fidelity end. The exact likelihood \eqref{eq:qchan} gains $1$ to $2$\,dB over decoding from the cell centers at 2 bits and less than $1$\,dB otherwise.

Two conventional readouts of equal payload delimit where random mixing pays off. Against per-pixel readout with fewer bits, CiP is $6.6$\,dB better at 1 bit per pixel and $3.1$\,dB better at 2, and the curves cross at 3, beyond which every per-pixel bit adds $6$\,dB while CiP saturates. Against $2\times2$ binning, however, CiP is $3.6$\,dB worse at 1 bit per pixel and $1.7$\,dB worse at 2. Natural images concentrate their energy at low spatial frequencies, which random weights do not exploit \cite{goyal}. Binning is itself a pattern of weights and selected pixels, so for natural scenes the weights should be structured, whereas random weights suit the pixel-sparse scenes considered next.

\subsection{Sparse Scenes}
For pixel-sparse scenes{\rev ,} we use GAMP and report the normalized mean-square error $\mathrm{NMSE}=\mathbb{E}\|\hat{\mathbf{x}}-\mathbf{x}\|^2/\mathbb{E}\|\mathbf{x}\|^2$, with ideal removal of the common mode, sweeping $\delta$ from $0.15$ to $0.9$ and the ADC resolution from 1 to 8 bits (curves not shown for space). For a block of $N=1024$ pixels, recovery sets in above $\delta\approx0.2$. Beyond this value, the unquantized error stays within $0.4$\,dB of the state evolution \cite{bayati} computed with \eqref{eq:sigeff} and falls only slowly with $\delta$, from $-25.6$ to $-34.3$\,dB between $\delta=0.25$ and $0.9$. At $\delta=0.5$, ADCs with 3, 4, 5, and 6 bits give $-17.7$, $-23.6$, $-27.4$, and $-30.0$\,dB against $-31.3$\,dB without quantization, so the variation floor is reached at about 6 bits, as for the rate in Fig.~\ref{fig:rate}. For a single column of $N=64$ pixels, which clips more often, the exact likelihood gains $2$ to $8$\,dB over treating quantization as noise. The cheapest way to reach $-20$\,dB is 5 bits at $\delta=0.25$, or $1.25$ bits per pixel, for $N=1024${\rev ,} and 4 bits at $\delta=0.45$ for one column. These operating points determine the edge energy, which we evaluated next.

\subsection{Edge Energy}
We finally estimate the energy of the node per pixel. A conventional node reads, converts with $b_c$ bits, and transmits each pixel. A CiP node performs $M$ readouts per block of $N$ pixels, each of which biases $N$ branches, rewrites $N$ weights, and yields one $b$-bit sample:
\begin{align}
    E_{\mathrm{conv}} &= E_{\mathrm{cell}} + F\,2^{b_c} + b_c E_b , \label{eq:econv}\\
    E_{\mathrm{CiP}} &= \delta\,\big[ N (E_{\mathrm{cell}}+E_w) + F\,2^{b} + b\,E_b \big]. \label{eq:ecip}
\end{align}
Here, $E_{\mathrm{cell}}=V_{\mathrm{DD}}\bar G\bar V T_{\mathrm{rd}}$ is the energy of one branch conducting during a readout window $T_{\mathrm{rd}}$, $E_w$ is the energy of one weight update, $F$ is the Walden figure of merit of the ADC \cite{murmann}, and $E_b$ is the radio energy per transmitted bit. We use $V_{\mathrm{DD}}=1$\,V, $\bar V=0.5$\,V, and $T_{\mathrm{rd}}=1\,\mu$s, so that $E_{\mathrm{cell}}=2.2$\,pJ, and we assume $E_w=1$\,pJ and $F=100$\,fJ per conversion step. The energy per bit $E_b$ is swept. Measured IEEE~802.15.4 and Bluetooth Low Energy (BLE) links spend about $0.42$ and $0.17\,\mu$J per delivered bit \cite{siekkinen}. For each $E_b$ and NMSE target, we select the pair $(\delta,b)$ that minimizes \eqref{eq:ecip}.

\begin{figure}[!t]
\centerline{\includegraphics[width=\plotwidth]{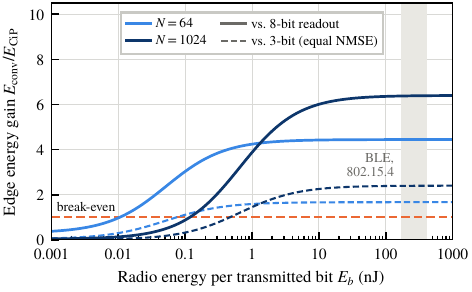}}
\caption{Edge energy gain of CiP readout for sparse scenes at a $-20$\,dB NMSE target versus the radio energy per bit, against a raw 8-bit conventional readout and a 3-bit one that meets the same NMSE.}
\vspace{-0.2cm}
\label{fig:energy}
\end{figure}

Fig.~\ref{fig:energy} shows the gain $E_{\mathrm{conv}}/E_{\mathrm{CiP}}$ for a $-20$\,dB target. For a BLE-class radio at $0.17\,\mu$J per bit, the radio takes more than $99.9\%$ of the CiP energy. Per pixel, mixing and weight updates cost $0.09$\,nJ, the ADC less than $1$\,pJ, and the radio $300$\,nJ. The gain therefore equals the payload ratio $b_c/(\delta b)$. Against a raw 8-bit readout, a single column saves a factor of $4.4$, which reduces the energy of a $64\times64$ frame from $5.5$ to $1.2$\,mJ, and a block of 1024 pixels, obtained by summing columns at the array output, saves a factor of $6.4$ at $(\delta,b)=(0.45,4)$ and $(0.25,5)$, respectively.

For sparse scenes, a conventional node that lowers its resolution to 3 bits also reaches $-20$\,dB, and against it the gains are $1.7$ and $2.4$. About a factor of two is thus specific to analog mixing, while the rest comes from matching the resolution to the fidelity. Because the overhead $\delta N(E_{\mathrm{cell}}+E_w)$ grows with the block length, CiP breaks even with the 3-bit readout only above $0.08$\,nJ per bit for one column and $0.5$\,nJ per bit for $N=1024$, which is two to three orders of magnitude below practical radios. The gain also shrinks as the target tightens because the variation floor pushes $\delta$ and $b$ up: at $-30$\,dB it is $2.2$ to $2.4$ against 8-bit readout and $1.4$ to $1.5$ against a 5-bit readout of equal fidelity. For the natural images of Fig.~\ref{fig:image}, 2 bits per pixel give a gain of $4$ against 8-bit readout and of $1.25$ to $1.5$ against a per-pixel readout of equal PSNR, whereas $2\times2$ binning needs half the payload of CiP.

\section{Conclusion}
We modeled the column bus of a FeFET CiP array as a multiple-access channel and its readout at the AP as a compressed-sensing problem. With reported MACPix parameters, the bus is limited by variation, not noise: a $2.4\%$ conductance spread fixes the signal-to-interference ratio at $27$\,dB, reduces the rate from $13.3$ to $4.5$ bits per readout, and makes a 5- to 6-bit ADC sufficient. Sparse scenes are recovered at $-20$\,dB from $1.25$ to $1.8$ bits per pixel, which lowers the edge energy by $4.4$ to $6.4$ times against raw 8-bit readout and $1.7$ to $2.4$ times against a conventional readout of equal fidelity; natural images reach $26.3$\,dB from 2 bits per pixel, above per-pixel readout but below $2\times2$ binning. The saving stems from compressing in the analog domain before the ADC, so that the radio carries fewer bits than there are pixels. Future work will address calibration of the conductance deviations at the AP, structured weights for natural images, and other CiP designs including phase-transition-material pixels.

\bibliographystyle{IEEEtran}
\bibliography{references2}

\end{document}